\documentclass[aps,twocolumn,preprintnumbers,superscriptaddress]{revtex4}

\usepackage{graphicx}  
\usepackage{subfigure}
\usepackage{multirow}

\usepackage{hyperref} 

\usepackage{fancyhdr}
\usepackage{longtable}
\usepackage{xcolor}
\usepackage{parskip}
\usepackage[T1]{fontenc}
\usepackage{dcolumn}   

\usepackage{bm}        
\usepackage{amsfonts}  
\usepackage{amsmath}   
\usepackage{amssymb}   
\usepackage{ulem}      
\usepackage{slashed}
\usepackage{braket}

\usepackage[T1]{fontenc}
\usepackage[utf8]{inputenc}
\usepackage{lmodern}

\usepackage[toc,page]{appendix}

\newtheorem{theorem}{Theorem}

\newenvironment{proof}{\textit{Proof.}}

\newenvironment{envCom}
  {\color{cyan}%
   \textbf{\small Comment: }\ignorespaces}
 {\smallskip}
 
 \newenvironment{envAdd}
  {\color{blue}%
   \textbf{\tiny [Add] }\ignorespaces}
 {\smallskip}
 
 \newenvironment{envRem}
  {\color{red}\ignorespaces}
 {\smallskip}

\begin{document}

\title{Maximal Secret Reconstruction, Teleportation and Bell's Inequality}


\author{Pratishtha Abrol}
\affiliation { Centre for Quantum Science and Technology, International Institute of Information Technology, Hyderabad, Hyderabad -- 500032, Telangana, India.\\ Center for Security, Theory and Algorithmic Research, International Institute of Information Technology, Hyderabad, Hyderabad -- 500032, Telangana, India.}
\author{Pahulpreet Singh}
\affiliation { Centre for Quantum Science and Technology, International Institute of Information Technology, Hyderabad, Hyderabad -- 500032, Telangana, India.\\ Center for Computational Natural Science and Bioinformatics, International Institute of Information Technology, Hyderabad, Hyderabad -- 500032, Telangana, India}
\author{Indranil Chakrabarty}
\affiliation { Centre for Quantum Science and Technology, International Institute of Information Technology, Hyderabad, Hyderabad -- 500032, Telangana, India.\\ Center for Security, Theory and Algorithmic Research, International Institute of Information Technology, Hyderabad, Hyderabad -- 500032, Telangana, India.}
\email{pratishtha.abrol@research.iiit.ac.in, pahulpreet.singh@research.iiit.ac.in,indranil.chakrabarty@iiit.ac.in}


\begin{abstract} 
A tripartite state is said to be a potential resource for secret sharing if the state imposes restrictions on the  teleportation fidelity of the bipartite dealer--reconstructor and dealer--assistant channels in addition of being useful for the state reconstruction \cite{singh2024controlled}. Given a secret shareable state in a pure three-qubit system \cite{acin2001three,acin2001classification}, we are able to characterize the set of states with maximum possible reconstruction fidelity (abbreviated as MSR states)  for a fixed value of the maximum teleportation fidelity that can be obtained out of both the dealer--receiver channels. 
Similarly for a value giving the maximum of Bell-CHSH value of both dealer--reconstructor and dealer--assistant channels, we are able to find the maximum achievable reconstruction fidelity. Interestingly, we find that all secret shareable states satisfy Bell's inequality in both dealer--reconstructor and dealer--assistant partitions. This brings out a new mutual exclusivity between secret shareable state and Bell's inequality violations. 
Our result  paves the way in identifying the best candidate among the secret sharing resource states in achieving the maximum reconstruction fidelity thus by setting the practical information transfer limit in a possible resource theoretic extension of secret sharing.  It also brings out a new kind of mutual exclusiveness between the bipartite correlation and in the ability of secret sharing in a tripartite setting.

\end{abstract}
\maketitle

\section{Introduction}
\noindent In quantum information theory, entanglement \cite{schrodinger1935discussion,horodecki2009quantum} plays a significant role in various communication protocols.
\noindent These involve sending of qubits (known or unknown) \cite{ horodecki1996teleportation, bennett1993teleporting, pati2023teleportation, chakrabarty2010teleportation, adhikari2008quantum, bennett2005remote} or classical bits \cite{bennett1992communication, srivastava2019one, das2015distributed, patro2017non, vempati2021witnessing, roy2018deterministic, wehner2018quantum} from one location to another without violating the no-cloning theorem \cite{wootters1982single}. Entanglement also enables the transfer of the coherence of quantum states using fewer resources \cite{pati2023teleportation}. 
All of these help us to develop resource theory of entanglement \cite{shahandeh2019resource}. Understanding its significance as a resource, entanglement in qubit-qubit \cite{datta2005entanglement, li2009generation}, qutrit-qutrit \cite{dajka2008origination, chkecinska2007separability} and qubit-qudit \cite{vinjanampathy2012quantum, mundra2019broadcasting} systems are broadcasted along with other quantifiers of resources. It can be further used to build up a quantum network\cite{das2018robust}.\\
\noindent Entanglement, as a form of nonlocal correlation plays a major role as a resource in sharing of classical secrets and also in the controlled reconstruction of quantum states.  Multiparty entanglement beyond bipartite entanglement acts as a resource in secure quantum communication tasks like controlled state reconstruction and secret sharing. In a general sense, secret sharing is a process where a dealer shares a secret with $n$ share holders such that only $k (k<n)$ or more share holders can reveal the secret but any number of share holders fewer than $k$ cannot reveal the secret. Such a secret sharing scheme is known as $(k,n)$ threshold scheme. This idea is further explored in quantum information theory where we are able to share classical bits as secrets using entangled resources. Interestingly, when we use a quantum state then the secret sharing protocol will have certain nuances. In the original article on quantum secret sharing by Hillery et. al. \cite{hillery1999quantum} this was referred as quantum information splitting. In consequent studies, the protocol has also been referred to as controlled quantum teleportation (CQT) \cite{garg2023estimation} and controlled state reconstruction (CSR)  and sometimes loosely as quantum secret sharing (QSS) \cite{hillery1999quantum, cleve1999share, karlsson1999quantum, bandyopadhyay2000teleportation, markham2008graph, li2010semiquantum, adhikari2010probabilistic, ray2016sequential, sazim2015retrieving, tittel2001experimental, schmid2005experimental, schmid2006experimental, bogdanski2008experimental}. In a recent work \cite{singh2024controlled} a new distinction was made in between the controlled state reconstruction (CSR) and quantum secret sharing (QSS). Based on that we are able to identify the conditions for a state to be useful for secret sharing within the set of states which are useful for controlled state reconstruction.\\
\noindent Following Bell's demonstration \cite{bell1964einstein}, in 1964, that quantum mechanics cannot be reconciled with the concept of local realism, there have been several advancements into investigating nonlocality in quantum systems. Bell inequalities, for example, establish an upper limit on the correlations observed in measurement statistics of multi party systems \cite{clauser1969proposed, svetlichny1987distinguishing}. Exceeding this upper limit implies the presence of a nonlocal correlation among the parties involved. For bipartite systems there is a more convincing way to quantify the Bell violation in the form of a closed formula \cite{horodecki1996teleportation}. This highlights the relation between the entanglement, bell's inequality and teleportation fidelity of states in a more explicit way with examples from either sides. Bipartite bell violation also plays a significant role in context of correlation and information processing abilities in multiparty situations \cite{pandya2016complementarity}.  \\ 
\noindent In this work, we begin with a three party framework where without any loss of generality we assume Alice is the dealer, Bob and Charlie are the assistant and reconstructor respectively. A three-qubit state will be called as a secret sharing resource if the state reconstruction fidelity at the Charlie's side is greater than $\frac{2}{3}$ in addition to the assumption that the fidelities of Alice-Charlie (Dealer--Reconstructor) and Alice-Bob (Dealer--Assistant) channels are less than $\frac{2}{3}$ \cite{singh2024controlled}. This comes from the idea that the individual information gain by the share holders should not be more than the classical limit. Here, in this work we study how the degree of faithfulness in transmission of a secret through a secret sharing protocol relates with the subsystem's individual ability to transmit a state through a teleportation protocol. We characterize those states which will have maximum reconstruction fidelity, given the maximum value of the bipartite teleportation fidelities of Alice-Charlie (Dealer--Reconstructor) and Alice-Bob (Dealer--Assistant) channels.  We also give a single parameter canonical representation of this maximal secret reconstructed resources (labeled as MSR state) from the set of all secret shareable resources in the pure tripartite state regime. We explore how the Bell-CHSH values of the bipartite  dealer--reconstructor, dealer--assistant subsystems correlate with the maximization of reconstruction fidelities. In particular we have shown that all secret shareable states satisfy Bell's inequality. Equivalently, we can say that three-qubit pure states with bipartite Bell violation is not suitable for quantum secret sharing.  This clearly brings out a new kind of mutually exclusiveness of bipartite Bell violation and the secret sharing capacity of the tripartite resource. In a larger sense, it brings out new kind of mutual exclusivity indicating that for processes like secret sharing we probably need tripartite states with minimum bipartite correlations. This entire study is carried out for the three-qubit pure states (analytical proofs are for three-qubit GHZ class of states with real coefficients and numerical results for the larger class of three  qubit GHZ states which  encompasses all possible three-qubit vectors \cite{acin2001three,acin2001classification}).  \\
The organization of the paper is explained as follows. In section II we give brief overview of the concepts which form the basis of our work. Section III introduces the theorem for maximal Secret reconstruction for given teleportation fidelities of the bipartite channels connecting the dealer. We also give the canonical representation of the states for which the maximal reconstruction is possible. In Section IV we discuss the correlation between the Bell's CHSH value of the same bipartite channels and the reconstruction fidelity of the state, In addition to that we  establish a mutual exclusiveness between the bipartite Bell violation and the secret shareable states. Section V is devoted to conclusions and scope for future discussions.

\section{Related concepts}
\noindent In this section, we provide an introduction to the various concepts that are employed in our study. These include teleportation fidelity, controlled state reconstruction fidelity and, the concept of bell's inequality violation. 



\subsection{Teleportation Fidelity} 
\noindent Quantum teleportation is the process of transferring unknown quantum information from one location to other. Teleportation fidelity is a measure of how well a quantum state can be faithfully transmitted using an entangled quantum channel aided with classical communication. The fidelity is a measure of the similarity between the teleported state and the original state. 
Let us consider a scenario involving Alice being the sender and Bob the receiver. To set up the protocol, Alice and Bob share an entangled resource. Suppose Alice wants to send the quantum state $\ket{\Psi}$. Alice would then perform a joint measurement on her particle of the shared entangled state and $\ket{\psi}$, based on which she sends classical information to Bob. Bob can then apply a unitary operation to retrieve the teleported state $\ket{\psi'}$. The teleportation fidelity can be given by the overlap between the original state and the teleported state:
\begin{equation}\label{eq:1}
    F^T = |\braket{\psi|\psi'}|^2
\end{equation}
Consecutively, the fidelity ranges from 0 to 1, 1 indicating perfect teleportation, as in the case of the Bell state $|\Psi\rangle=\frac{1}{\sqrt{2}}(|00\rangle+|11\rangle )$. We can transfer an unknown state perfectly \cite{hillery1999quantum}, considering no interference from external sources. In other words, Bell state can be considered as an optimal resource for transferring quantum information.
\noindent It was also shown that coherence of a quantum state can also be teleported with much lesser resource \cite{pati2023teleportation}. A fidelity less than 1 indicates some deviation or error in the teleportation protocol. 

\noindent It can be shown that one can still carry out teleportation without having an entangled resource, classically to achieve a fidelity $\frac{2}{3}$, if not $1$. Any quantum resource giving a fidelity greater than $\frac{2}{3}$, is in general considered as a resource as there is quantum advantage associated in beating the classical limit. As a resource, we consider a two-qubit mixed state  \cite{horodecki1996teleportation},
\begin{equation}\label{eq:2}
\rho_{AB}=\frac{1}{4}(I \otimes I + \sum\limits_{i} r_i.\sigma_{i} \otimes I+ \sum\limits_{i} s_{i}.I \otimes \sigma_{i} + \sum\limits_{ij} t_{ij} \sigma_{i} \otimes \sigma_{j}),   
\end{equation}
where $\sigma_{i} = (\sigma_{1}, \sigma_{2}, \sigma_{3})$ are the Pauli matrices; $r_{i} = (r_{1}, r_{2}, r_{3})$, $s_{i} = (s_{1}, s_{2}, s_{3})$ are the local Bloch vectors and $t_{ij}=Tr(\rho( \sigma_{i} \otimes \sigma_{j} ))$ are the elements of the correlation matrix T = $[t_{ij}]_{3\times3}$. The maximum teleportation fidelity that can be achieved for such a state $\rho_{AB}$ is given by,
\begin{equation}\label{eq:3}
F_{AB}=\frac{1}{2}(1+\frac{1}{3}\vartheta_2(\rho)),    
\end{equation}
where $\vartheta_2(\rho)=Tr(\sqrt{T^{\dagger}T})$. The above formula characterizes the teleportation fidelity in terms of the parameters of the resource state. For a teleportation protocol to be successful, the teleportation fidelity should be greater  than $\frac{2}{3}$. This does not mean that there do not exist entangled states for which the teleportation fidelity is less than $\frac{2}{3}$ \cite{horodecki1996teleportation,adhikari2008quantum,chakrabarty2011deletion,chakrabarty2010teleportation}. In other words, not all entangled states guarantee a successful teleportation. 

\subsection{Controlled State Reconstruction and Secret Sharing} 
\noindent  A controlled state reconstruction process involves reconstructing a quantum state shared by the dealer at the share holder's location. This requires cooperation of other shareholder in the process of reconstruction. In the most simplistic scenario when there is one dealer (Alice) and two share holders (Bob and Charlie), we consider a general three-qubit state,
\begin{eqnarray}\label{eq:4}
 &&\rho_{ABC}= 
 \frac{1}{8}[I^{\otimes 3} +\sum_{i=1}^3 a_i.\sigma _i\otimes I^{\otimes 2}+ {}\nonumber\\&&
 \sum_{j=1}^3 I \otimes b_j.\sigma_j \otimes I + \sum_{k=1}^3 I^{\otimes 2} \otimes c_k.\sigma_k+ {}\nonumber\\&&
 \sum_{i,j=1}^3 q_{ij} \sigma_i \otimes \sigma_j \otimes I + \sum_{i,k=1}^3 r_{ik}\sigma_i \otimes I \otimes \sigma_k + {}\nonumber\\&&\sum_{j,k=1}^3 s_{jk} I\otimes \sigma_j \otimes \sigma_k +\sum_{i,j,k=1}^3 t_{ijk} \sigma_i \otimes \sigma_j \otimes \sigma_k
 ],   \label{eq:resource_state}
\end{eqnarray}

\noindent as the resource state. Here $a_i,b_j,c_k$ are local Bloch vectors and the correlation matrices are given by,
$Q=\{q_{ij}\}=Tr(\rho_{ABC}( \sigma_i \otimes \sigma_j \otimes I))$, $R=\{r_{ik}\}=Tr(\rho_{ABC}( \sigma_i \otimes I \otimes \sigma_k))$ and $S=\{s_{jk}\}=Tr(\rho_{ABC}( I \otimes \sigma_j \otimes \sigma_k))$. These correlation matrices are of order $3 \times 3$. Here $\tau=t_{ijk}=Tr(\rho_{ABC}( \sigma_i \otimes \sigma_j \otimes \sigma_k))$ is the correlation tensor. The expression for maximum reconstruction fidelity for a three-qubit state $\rho_{ABC}$ when Alice is the dealer, Bob is the assistant and Charlie is the reconstructor, is given by
\begin{equation}\label{eq:5}
    \mathcal{F}_{ABC}^{CSR} = \frac{1}{2} + \frac{1}{6} \vartheta_3(\rho)
\end{equation}
\noindent where $\vartheta_3(\rho)$ is a function of the parameters of the state, given by 
\begin{equation}\label{eq:6}
\vartheta_3(\rho) = \frac{\|R + T\|_1 + \|R - T\|_1}{2}.    
\end{equation}

\noindent If the reconstruction fidelity $\mathcal{F}^{CSR}>\frac{2}{3}$ , then we have a quantum advantage as mentioned in \cite{singh2024controlled}. This measure helps us to identify the states that are useful for controlled state reconstruction.\\  

\noindent Note that the secret sharing process is not exactly equivalent to a reconstruction process. That is, along with successful reconstruction with a fidelity greater than $\frac{2}{3}$, we have to ensure that the share holders will not be able to reconstruct the secret on their own. This independent reconstruction can be taken equivalent to a teleportation protocol and only be possible if the bipartite channel achieve teleportation fidelity more than $\frac{2}{3}$.  So to identify a three-qubit state $\rho_{ABC}$ which is useful for the secret sharing protocol we need additional constraints that inhibit teleportation across bipartitions. Since, for a channel to be useful for teleportation, we have $F > \frac{2}{3}$, we have the conditions:
\begin{equation}\label{eq:7}
F_{AB} \leq\frac{2}{3}, F_{AC} \leq\frac{2}{3}    
\end{equation}
\noindent along with the condition $\mathcal{F}^{CSR}>\frac{2}{3}$ as shown in \cite{singh2024controlled} for a dealer ($A$) and share holders ($B$ and $C$) setting. These conditions together ensure faithful transport of the secret, while also accounting for dishonest recipients.\\

\noindent \textbf{Note:} 1. In the later section we will call the maximum achievable teleportation fidelity of a channel as simple teleportation fidelity to avoid confusion.\\
2. From now on, for the purpose of this paper, we will be referring to teleportation fidelity in the notation $F$ and reconstruction fidelity as $\mathcal{F}$.

\subsection{Bell's Inequality Violation}
\noindent Bell's inequality violation is a phenomenon in quantum mechanics that demonstrates the non classical nature of correlations between entangled particles. The violation of the inequality implies the presence of nonlocal correlations that cannot be explained by classical realism. According to local realism, physical properties of a system have well defined values, independent of measurement settings on distant particles. The Bell-CHSH inequality, for a pair of entangled pparticles with measurement settings $a,b$ and outcomes $A,B$ is given by:
\begin{equation}\label{eq:8}
    |\mathcal{S}| \leq 2
\end{equation}
where $\mathcal{S}$ is the CHSH correlation given by $\mathcal{S} = E(a,b) + E(a',b) + E(a,b') - E(a',b')$ and $E(a,b)$ is the correlation between the outcomes $A$ and $B$ for measurement settings $a$ and $b$.
All bipartite pure entangled states violate the Bell inequality \cite{yu2012all}. For a two-qubit mixed state to violate the Bell-CHSH inequality, it must hold the condition that the Bell-CHSH value $\mathcal{S}$ is greater than 2, to imply the presence of nonlocal correlation that cannot be described by the assumptions of local realism. The Bell-CHSH observable quantity for any arbitrary bipartite state $\rho$ (\ref{eq:2}) can be defined as 
\begin{equation}\label{eq:9}
    \mathcal{S}(\rho) = 2 \sqrt{M(\rho)}
\end{equation}
where $M(\rho) = m_1 + m_2$. $m_1$ and $m_2$ are the two greatest eigenvalues of $T^T_{\rho}T_{\rho}$, the correlation matrix for the two-qubit state being $T_{\rho}$ defined as $(T_{\rho})_{ij} = Tr[\rho (\sigma_i \otimes \sigma_j )]$ where $\sigma_i$ represent the respective Pauli Matrices. Drawing from the condition for Bell Inequality Violation, it is clear that, $M(\rho) \leq 1$ is equivalent to the Bell inequality not being violated.\\ 

\noindent It is interesting to note, that in the context of teleportation fidelity, it is an established fact that $\vartheta_2 (\rho) \geq M(\rho)$ \cite{horodecki1996teleportation}, for an arbitrary mixed spin-$\frac{1}{2}$ state to be useful for teleportation.
\noindent This inequality results from the definition of the quantity $\vartheta_2 (\rho) = \Sigma_{i=1}^3 \sqrt{m_i}$ with $m_i$ being the eigenvalues of the matrix $T^T T$. Since $M(\rho) > 1$ for a state violating Bell's inequality, we can estimate 
\begin{eqnarray}\label{eq:10}
    F_{max} &\geq& \frac{1}{2}(1 + \frac{1}{3}M(\rho)) > \frac{2}{3}
\end{eqnarray}

\noindent Note that the absence of Bell's inequality violation does not rule out the presence of entanglement or nonlocality.

\section{ Reconstruction Fidelity Versus Teleportation Fidelity in Secret Sharing}

\noindent In this section we obtain the relation between the reconstruction fidelity of the shared pure tripartite  resource state and the maximum teleportation fidelity between the two bipartite channels involving the dealer. In particular, we consider those states that are useful for secret sharing in the class of the generalized GHZ (Greenberger-Horne-Zeilinger) states \cite{acin2001three,acin2001classification} representing  all possible pure states in the tripartite scenario. We show that there exists a limit to the maximal reconstruction fidelity such a tripartite state can achieve, given the maximal achievable  teleportation fidelity that can be obtained  from the bipartite channels (dealer--reconstructor/dealer--assistant). Furthermore, we are also able to identify the family of states that will achieve maximum reconstruction fidelity by giving its canonical form.\\ 
\noindent Let us choose a state from the generalized GHZ (Greenberger-Horne-Zeilinger) class of states with real coefficients shared between three parties -- Alice ($A$) as the dealer, Bob ($B$) as the assistant and Charlie ($C$) as the reconstructor. It is important to mention that these set of states that can be converted into the GHZ state $\frac{1}{\sqrt{2}} (\ket{000} + \ket{111})$ using Stochastic Local Operation and Classical Communication (SLOCC) with non zero probability \cite{acin2001three,acin2001classification}. These states can be expressed as :
\begin{eqnarray}\label{eq:12}
    \ket{\psi(GHZ^R)}_{ABC} =&& \lambda_0 \ket{000} + \lambda_1 \ket{100} + \lambda_2 \ket{101}{}\nonumber\\&& + \lambda_3 \ket{110} + \lambda_4 \ket{111},
\end{eqnarray}
where $\lambda_i \geq 0, \Sigma_i \lambda_i ^2 = 1$  
This state is a derivative  of the Acin state \cite{acin2001classification} obtained by putting the complex part $\phi$ associated with $\lambda_1$ as $e^{i\phi}\lambda_2$ to $0$. The reconstruction fidelity at Charlie's ($C$) location, for such a state in terms of $\lambda_i$, is given by,
\begin{eqnarray}\label{eq:13}
&&\mathcal{F(GHZ^R)}_{ABC} = \frac{1}{2} + \frac{1}{6} \vartheta_3(\rho);{}\nonumber\\&&
\vartheta_3 (\rho) = 4 \lambda_0 \text{max} \{\lambda_2, \lambda_4\} + 1
\end{eqnarray}
\noindent Here in this setting we consider the maximum ($F_{max}$) of the teleportation fidelities $F_{AB}$ and $F_{AC}$ of both dealer--assistant ($A-B$) and dealer--reconstructor ($A-C$) channels of the tripartite  state $\ket{\psi(GHZ^R)}$. In other words, $F_{max}=max(F_{AB},F_{AC})$. In the same spirit we can write $\vartheta_2 (\rho)$ of $F_{max}$ as $\vartheta_2 (\rho)=max(\vartheta_2 (\rho_{AB}),\vartheta_2 (\rho_{AC}))$. It turns out that  $\vartheta_2(\rho)$  depends upon two correlation matrices $R_{AC}$ and $Q_{AB}$ of the channels $AC$ and $AB$ respectively:
\begin{equation}\label{eq:15}
\vartheta_2 (\rho) = \text{max} \{||R||_{AC}, ||Q||_{AB}\} , 
\end{equation}
where
\begin{eqnarray}\label{eq:16}
    &&||R|| = 2 \lambda_0 \lambda_2 + 2 \sqrt{ \lambda_0^2 \lambda_2^2 + (\lambda_1 \lambda_2 + \lambda_3 \lambda_4)^2 } {}\nonumber\\&&
    + \sqrt{4 \lambda_0^2 \lambda_1^2 + (\lambda_0^2 - \lambda_1^2 + \lambda_4^2 +\lambda_2^2 -\lambda_3^2)^2} 
\end{eqnarray}
\begin{eqnarray}\label{eq:17}
    &&||Q|| = 2 \lambda_0 \lambda_3 + 2 \sqrt{ \lambda_0^2 \lambda_3^2 + (\lambda_1 \lambda_3 + \lambda_2 \lambda_4)^2 } {}\nonumber\\&&
    + \sqrt{4 \lambda_0^2 \lambda_1^2 + (\lambda_0^2 - \lambda_1^2 + \lambda_4^2 +\lambda_3^2 -\lambda_2^2)^2} 
\end{eqnarray}
\noindent It can be seen from the equations \eqref{eq:16} and \eqref{eq:17} that the value of $\vartheta_2(\rho)$ depends on the value of $\lambda_2$ and $\lambda_3$. If $\lambda_2 > \lambda_3$ we have $\vartheta_2(\rho)= ||R||_{AB}$ and if $\lambda_3 > \lambda_2, \vartheta_2(\rho)= ||Q||$. At $\lambda_2 = \lambda_3$ the values of $||R|| = ||Q||$. \\

\begin{theorem}\label{th:1}
    Given a three-qubit pure state $\ket{\psi(GHZ^R)}_{ABC}$, which is achieving a quantum advantage for secret sharing, implying that $F_{AB} \leq \frac{2}{3}$  and $F_{AC} \leq \frac{2}{3}$ ( equivalently we can say that $F_{max} \leq \frac{2}{3}$ or $\vartheta_2 (\rho) \leq 1$) and $\mathcal{F} _{ABC} > \frac{2}{3}$ (or equivalently,$\vartheta_3 (\rho) > 1$), for a teleportation fidelity of $F_{max}$, the reconstruction fidelity is bounded by the following inequality:
    \begin{equation}\label{eq:18}
        \mathcal{F_{ABC}} \leq 2 F_{max} - \frac{1}{3}
    \end{equation}
    which can also be expressed as
    \begin{equation}\label{eq:19}
        \vartheta_3 (\rho) \leq 1 + 2 \vartheta_2 (\rho)
    \end{equation}
\end{theorem}
\vspace{0.2cm}
\begin{proof}   Let us define 2 states, an arbitrary state $\rho^a \equiv \{\lambda_0,\lambda_1,\lambda_2,\lambda_3,\lambda_4\}$ and a boundary state that achieves equality in the above theorem, $\rho^b \equiv \{\Lambda_0,\Lambda_1,\Lambda_2,\Lambda_3,\Lambda_4\}$. We can divide this proof into three parts as
\begin{equation}
    \vartheta_3(\rho^a) \overset{(1)}{\leq} \vartheta_3(\rho^b) \overset{(2)}{=} 2 \vartheta_2(\rho^b) + 1 \overset{(3)}{\leq} 2 \vartheta_2(\rho^a)+1
\end{equation}
with
\begin{enumerate}
    \item $\forall \rho^a \exists \rho^b: \vartheta_3(\rho^a) \leq \vartheta_3(\rho^b)$, given $\vartheta_2(\rho^a) =\vartheta_2(\rho^b)$
    \item $\forall \rho^b; \vartheta_3(\rho^b) = 2\vartheta_2(\rho^b) + 1$
    \item $\forall \rho^a \exists \rho^b: \vartheta_2(\rho^a) \geq \vartheta_2(\rho^b)$
\end{enumerate}
From the definition of teleportation fidelity \eqref{eq:15} it is evident that, keeping $\lambda_0, \lambda_1, \lambda_4$ constant, increasing $\lambda_2$ decreases $\lambda_3$ and hence, increases $||R||$ and decreases $||Q||$ and vice versa. To minimize $\vartheta_2(\rho)$ we require $\lambda_2= \lambda_3= 0$. Hence we can define $\rho^b \equiv \{\Lambda_0,\Lambda_1,0,0,\Lambda_4\}$, at which point 
\begin{equation}
    \vartheta_2(\rho^b) = \sqrt{1 + 4\lambda_4^4 + 4\lambda_4^2\lambda_0^2 - 4 \lambda_4^2} \leq \vartheta_2(\rho^a)
\end{equation}
Thus, part 3 is true.
Now, for $\rho^b$ we have $\vartheta_3(\rho^b) = 4\lambda_0\lambda_4 + 1$. Replacing these values in the theorem, we can achieve equality at the point $\lambda_4 = \frac{1}{\sqrt{2}}$. Since this establishes part 2 and the existence of a boundary, we can define this state as  $\rho^b \equiv \{\frac{1}{\sqrt{2}} \cos\theta, \frac{1}{\sqrt{2}} \sin\theta, 0, 0, \frac{1}{\sqrt{2}}\}$. We can now calculate the values for $\vartheta_2(\rho^b)$ and $\vartheta_3(\rho^b)$ in terms of $\theta$ as $\vartheta_2(\rho^b)=\cos\theta$ and $\vartheta_3(\rho^b) = 2\cos\theta + 1$. \\
\noindent Now,, assert that $\vartheta_2(\rho^a) = \vartheta_2(\rho^b) = \cos\theta$. \\

\noindent\textbf{CASE 1: If $\lambda_2>\lambda_3 :$} 
\begin{equation}
    \vartheta_2(\rho^a) = ||R_a|| = \cos\theta =\vartheta_2(\rho^b)
\end{equation}
\begin{enumerate}
    \item \textbf{if $\lambda_2 \geq \lambda_4 \Rightarrow \vartheta_3(\rho^a) = 4\lambda_0\lambda_2+1$} \\
    Replace the value of $\cos\theta$ in $\vartheta_3(\rho^b)$ 
    \begin{align*}
        \vartheta_3(\rho^b) &=4 \lambda_0 \lambda_2 + 4 \sqrt{\lambda_0^2 \lambda_2^2 + (\lambda_1\lambda_2 + \lambda_3 \lambda_4)^2} \\ 
&+ 2\sqrt{4 \lambda_0^2 \lambda_1^2 + (\lambda_0^2 - \lambda_1^2 + \lambda_4^2 +\lambda_2^2 - \lambda_3^2)^2} + 1 \\ 
& > 4\lambda_0\lambda_2 + 1 =  \vartheta_3(\rho^a)
    \end{align*}
    \item \textbf{if $\lambda_4 > \lambda_2 \Rightarrow \vartheta_3(\rho^a) = 4\lambda_0\lambda_4+1$} \\ 
    Simplifying the last term it is obvious that,
    \begin{align*}
        \vartheta_3(\rho^b) &>4 \lambda_0 \lambda_2 + 4 \sqrt{\lambda_0^2 \lambda_2^2 + (\lambda_1\lambda_2 + \lambda_3 \lambda_4)^2} \\ 
&+ 2\sqrt{4\lambda_0^2\lambda_4^2} + 1 \\ 
&> 4\lambda_0\lambda_4 + 1 =  \vartheta_3(\rho^a)
    \end{align*}
\end{enumerate}
Thus, if $\lambda_2>\lambda_3$ and $\vartheta_2(\rho^a) = \vartheta_2(\rho^b)$ then $\vartheta_3(\rho^b) > \vartheta_3(\rho^a)$. \\

\noindent\textbf{CASE 2: If $\lambda_2<\lambda_3 :$} 
\begin{equation}
    \vartheta_2(\rho^a) = ||Q_a|| = \cos\theta =\vartheta_2(\rho^b)
\end{equation}
\begin{enumerate}
    \item \textbf{if $\lambda_3 > \lambda_2 \geq \lambda_4 \Rightarrow \vartheta_3(\rho^a) = 4\lambda_0\lambda_2+1$} \\
    Replace the value of $\cos\theta$ in $\vartheta_3(\rho^b)$ 
    \begin{align*}
        \vartheta_3(\rho^b) &>4 \lambda_0 \lambda_3 + 1 \\ 
& > 4\lambda_0\lambda_2 + 1 =  \vartheta_3(\rho^a)
    \end{align*}
    \item \textbf{if $\lambda_4 > \lambda_2 \Rightarrow \vartheta_3(\rho^a) = 4\lambda_0\lambda_4+1$} \\ 
    Simplify the last term to the form:
    \begin{align*}
        \vartheta_3(\rho^b) &>4 \lambda_0 \lambda_3 + 4 \sqrt{\lambda_0^2 \lambda_3^2 + (\lambda_1\lambda_3 + \lambda_2 \lambda_4)^2} \\ 
&+ 2\sqrt{4\lambda_0^2\lambda_4^2 } + 1 \\ 
& > 4\lambda_0\lambda_4 + 1 =  \vartheta_3(\rho^a)
    \end{align*}
\end{enumerate}
Thus, if $\lambda_3>\lambda_2$ and $\hspace{0.2cm} \vartheta_2(\rho^a) = \vartheta_2(\rho^b)$ then $\vartheta_3(\rho^b) > \vartheta_3(\rho^a)$. Thus, for all arbitrary states belonging to the GHZ class as defined above, the theorem is true, achieving equality only if $\lambda_2=\lambda_3=0$. $\blacksquare$
\end{proof}
\begin{figure}[h]
    \centering
    \includegraphics[width=0.4\textwidth]{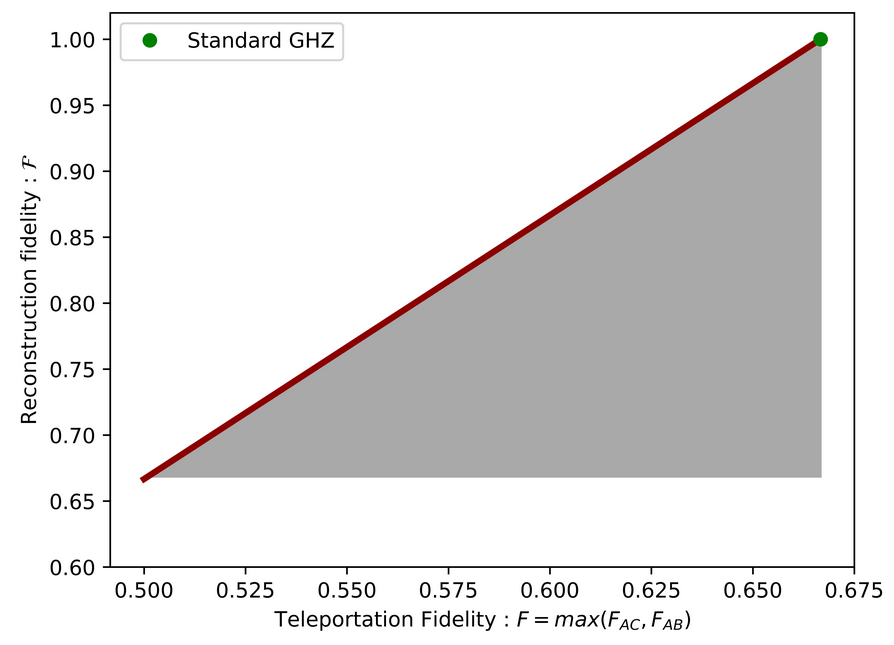}
    \caption{Fidelity of Reconstruction vs Teleportation fidelity: The reconstruction fidelity of the tripartite pure state \cite{acin2001three,acin2001classification} is plotted against the maximum of the teleportation fidelities of the (dealer--reconstructor) and (dealer--assistant) channels. The red coloured line indicates the MSR states with the standard GHZ state shown as the green point is considered as an example to this class.}
    \label{fig:rf-vs-tf}
\end{figure}

\noindent Thus, we can establish the fact that there exists a limit to the maximal reconstruction fidelity $\ket{\psi(GHZ^R)}_{ABC}$ can achieve, given the  teleportation fidelity of the bipartite reduced states of $AB$ and $AC$ channels. Interestingly, in the the Figure (\ref{fig:rf-vs-tf}), we show numerically that aour result holds for Generalized GHZ class of states (sometimes referred as Acin states) : $ \ket{\psi(GHZ^R)}_{ABC} = \lambda_0 \ket{000} + \lambda_1 e^{i\phi}\ket{100} + \lambda_2 \ket{101}+ \lambda_3 \ket{110} + \lambda_4 \ket{111}$ cannonically representing any three-qubit vector in general.\\

\noindent \textit{Maximal Secret Reconstructed State (MSR) :} In this article we are able to characterize those states for which maximum secret reconstruction is possible among the secret sharing resources. We call these states to be Maximal Secret Reconstructed State (MSR). For a given value of the maximum teleportation fidelity $F_{max}$, these states give the maximum reconstruction fidelity. 
This class of states are obtained from the equality condition in Theorem \ref{th:1}, by putting the values $\lambda_2 = \lambda_3 = 0$ and $\lambda_4 = \frac{1}{\sqrt{2}}$ in the state $\ket{\psi(GHZ^R)}$. This can also be identified in the Figure (\ref{fig:rf-vs-tf} by the red line over the ash triangle. States belonging to this family can be represented as:
\begin{equation}\label{eq:27}
    \ket{\psi}_{GHZ^R} = \lambda_0 \ket{000} + \lambda_1 \ket{100} + \frac{1}{\sqrt{2}} \ket{111} 
\end{equation}
where, $\lambda_i \in \mathbf{R}, \mu_i \geq 0, \Sigma_{i} \lambda_i^2 = \frac{1}{2}$. Canonically, it can also have a single parameter representation as:
\begin{equation}\label{eq:28}
    \ket{\psi}_{GHZ^R} = \frac{1}{\sqrt{2}} (\cos{\theta} \ket{000} + \sin{\theta} \ket{100}) + \frac{1}{\sqrt{2}} \ket{111} 
\end{equation}
with $\theta \in [0, \frac{\pi}{2}]$.\\

\noindent \textit{Example 1. The GHZ state: }
A classic example of a state belonging to this class is the GHZ state ($\theta = \frac{\pi}{2}$): $\ket{GHZ} = \frac{1}{\sqrt{2}} \ket{000} + \frac{1}{\sqrt{2}} \ket{111}$, where $\lambda_0 = \frac{1}{\sqrt{2}}$ and $\lambda_1 = 0$.  For this state, we find that the maximum teleportation fidelity of the bipartite channels $F_{max} (GHZ) = \frac{2}{3}$ and the reconstruction fidelity $\mathcal{F(GHZ)} = 1$ which satisfies the equation \ref{eq:18}.
This clearly indicates that the GHZ state falls into the class of Maximum secret reconstructed class of state (MSR) by achieving maximum possible reconstruction fidelity in addition of being classified as a secret sharing resource (labeled as the green dot in the Figure (\ref{fig:rf-vs-tf}).\\

\noindent \textit{Example 2. $\theta = \frac{\pi}{4}$: }
Such a state would take the form $\ket{\psi} = \frac{1}{2} \ket{000} + \frac{1}{2} \ket{100} + \frac{1}{\sqrt{2}}\ket{111}$. Knowing the coefficients, we can substitute them in the definitions for bipartite teleportation fidelity and reconstruction fidelity. For teleportation fidelity, from \eqref{eq:15}, $\vartheta_2 (\rho) = \frac{1}{\sqrt{2}}$, and hence $F_{max} = \frac{6+\sqrt{2}}{12}$ $< \frac{2}{3}$ \eqref{eq:3}. From definition of reconstruction fidelity, $\vartheta_3(\rho) = \sqrt{2}+1$, and from \eqref{eq:5} $\mathcal{F}= \frac{4+\sqrt{2}}{6}$ $> \frac{2}{3}$. Clearly, $\mathcal{F} - 2F = - \frac{1}{3}$ which achieves equality on Theorem \ref{th:1}. Clearly the state $|\psi\rangle$ falls into the MSR class of states. 

\section{Reconstruction fidelity versus Bell's Inequality Violation}

\noindent In this section, we explore the relationship between the Bell's Bell-CHSH value of the dealer-reconstructor and dealer-assistant channels with the reconstruction fidelity of the tripartite GHZ class state. \\
\noindent Considering the generalized GHZ class of states defined in equation \eqref{eq:12}, we find the reduced density matrices $\rho_{AC}$ and $\rho_{AB}$ for the dealer--reconstructor and dealer--assistant bipartitions respectively. It can be calculated that the correlation matrix for each bipartition, represented as $T_{\rho} = Tr (\sigma_i \otimes \sigma_j \rho)$ which takes the value of $R$ for the subsystem AC and $Q$ for subsystem AB. Similar to the teleportation fidelity, the eigenvalues of the matrix $T_{\rho}^T T_{\rho}$ will differ by the factor of $\lambda_2$ and $\lambda_3$ for the two bipartitions. Note that all three of the eigenvalues from each bipartition will remain non-negative. \\
 
\begin{theorem}\label{c:1.2}
    (Mutual Exclusivity Theorem: ) All states that are useful for secret sharing satisfies Bell's inequality ($\mathcal{S}_{max} \leq 2$). Equivalently, the maximum  bipartite Bell's Inequality Violation ($\mathcal{S}_{max}>2$) would imply that the state is not useful for faithful secret sharing.
\end{theorem}

\noindent \begin{proof}
We already know that for a state to be useful for secure secret sharing, we have the 2 necessary conditions; one for the bipartitions, ( $F_{AB} \leq \frac{2}{3}$ and $F_{AC} \leq \frac{2}{3}$ or equivalently $F_{max} \leq \frac{2}{3}$) and for the entire three-qubit state $\mathcal{F}\geq \frac{2}{3}$.  We can use this condition to establish a similar limit on Bell-CHSH value ($\mathcal{S}_{max}=max\{\mathcal{S}_{AC},\mathcal{S}_{AB}\}$ or equivalently $M_{max} = \max\{M(\rho_{AC}), M(\rho_{AB})\}$) through equation \eqref{eq:10}.
\begin{eqnarray}
&&F_{max} \leq \frac{2}{3}\Rightarrow \frac{2}{3} \geq \frac{1}{2} (1 + \frac{1}{3} M_{max}){}\nonumber\\&&
\Rightarrow 1 \geq M_{max} \Rightarrow \mathcal{S}_{max} \leq 2
\end{eqnarray} 
\noindent This clearly indicates that those states for which $F_{max} \leq \frac{2}{3}$  are not nonlocal in the dealer--reconstructor and dealer--assistant bipartite regime as it is independent of whether $\mathcal{F} \geq \frac{2}{3}$. So eventually for secret shareable states  being a subset of that will satisfy the same conclusion.  This makes sense from the perspective that if there are more correlation in the bipartite channels more information individual can access without involving others. This creates a mutual exclusiveness  of the secret sharing capacity and the bipartite nonlocality.   $\blacksquare$
\end{proof} 

\begin{theorem}\label{th:2}
    Given a three-qubit pure state $\ket{\psi(GHZ^R)}$ which is useable for secret sharing under the below mentioned conditions:
    \begin{align*}
        F_{max} \leq \frac{2}{3} \Rightarrow \vartheta_2(\rho) \leq 1 \\
        \mathcal{F}_{ABC} \geq \frac{2}{3} \Rightarrow \vartheta_3(\rho) \geq 1
    \end{align*}
    the relation between Bell's CHSH value $\mathcal{S}_{max}=max(\mathcal{S}_{AB}, \mathcal{S}_{AC})$ (equivalently $M_{max}=max(M_{AB}, M_{AC})$) and reconstruction fidelity $\mathcal{F}$ can be presented as:
    \begin{equation}\label{eq:30}
        6 \mathcal{F}_{ABC} \leq \mathcal{S}_{max} +4
    \end{equation}
    or, in terms of their factors as:
    \begin{equation}\label{eq:31}
        \vartheta_3 (\rho) \leq 2\sqrt{M_{max}} +1
    \end{equation}
    
\end{theorem}

\noindent \begin{proof} As before, we define 2 states, an arbitrary state $\rho^a \equiv \{\lambda_0,\lambda_1,\lambda_2,\lambda_3,\lambda_4\}$ and a boundary state, $\rho^b \equiv \{\Lambda_0,\Lambda_1,\Lambda_2,\Lambda_3,\Lambda_4\}$. We can divide this proof into three parts as
\begin{equation}
    \vartheta_3(\rho^a) \overset{(1)}{\leq} \vartheta_3(\rho^b) \overset{(2)}{=} 2 \sqrt{M_{\max}(\rho^b)} + 1 \overset{(3)}{\leq} 2 \sqrt{M_{\max}(\rho^a)} +1
\end{equation}
with
\begin{enumerate}
    \item $\forall \rho^a \exists \rho^b: \vartheta_3(\rho^a) \leq \vartheta_3(\rho^b)$, given $M(\rho^a) =M(\rho^b)$
    \item $\forall \rho^b; \vartheta_3(\rho^b) = 2\sqrt{M(\rho^b)} + 1$
    \item $\forall \rho^a \exists \rho^b: M(\rho^a) \geq \vartheta_2(\rho^b)$
\end{enumerate}
We have already seen that $M(\rho)$ is minimized at $\lambda_2=\lambda_3=0$, thus part $3$ is true at $\rho^b \equiv \{\Lambda_0,\Lambda_1,0,0,\Lambda_4\}$.
Following similar steps as in proof for Theorem 1, we can obtain the equality at $\lambda_4=\frac{1}{\sqrt{2}}$ thus defining the family of states lying on the boundary in the form $\rho^b \equiv \{\frac{1}{\sqrt{2}} \cos\theta, \frac{1}{\sqrt{2}} \sin\theta, 0, 0, \frac{1}{\sqrt{2}}\}$. Again, we have, in terms of $\theta$, $M(\rho^b) = \cos^2\theta$ and $\vartheta_3(\rho^b) = 2\cos\theta + 1$. \\

\noindent\textbf{CASE 1: If $\lambda_2>\lambda_3 :$} 
\begin{eqnarray}
    M(\rho^a) &&= 1 - 4 (\lambda_1\lambda_4 + \lambda_2\lambda_3)^2 + 4\lambda_0^2 (\lambda_2^2 - \lambda_3^2) {}\nonumber\\&&
    = \cos^2\theta = M(\rho^b)
\end{eqnarray}
\begin{enumerate}
    \item \textbf{if $\lambda_2 \geq \lambda_4 \Rightarrow \vartheta_3(\rho^a) = 4\lambda_0\lambda_2+1$} \\
    Replace the value of $\cos\theta$ in $\vartheta_3(\rho^b)$ 
    \begin{align*}
    &\vartheta_3(\rho^b) = 2\cos\theta +1 \\
    &= 1+ 2\sqrt{1 - 4(\lambda_1\lambda_4 + \lambda_2\lambda_3)^2 + 4\lambda_0^2 (\lambda_2^2 - \lambda_3^2)} \\
    & > 1 + 2\sqrt{ 4\lambda_0^2\lambda_3^2} > 1 + 4\lambda_0\lambda_2 = \vartheta_3(\rho^a)
    \end{align*}
    \item \textbf{if $\lambda_4 > \lambda_2 \Rightarrow \vartheta_3(\rho^a) = 4\lambda_0\lambda_4+1$} \\ 
    Similarly, simplify the last term to the form:
    \begin{align*}
    &\vartheta_3(\rho^b) = 2\cos\theta +1 \\
    &= 1+ 2\sqrt{1 - 4(\lambda_1\lambda_4 + \lambda_2\lambda_3)^2 + 4\lambda_0^2 (\lambda_2^2 - \lambda_3^2)} \\
    & > 1 + 2\sqrt{ 4\lambda_0^2\lambda_4^2} = 1 + 4\lambda_0\lambda_4 = \vartheta_3(\rho^a)
    \end{align*}
\end{enumerate}
Thus, if $\lambda_2>\lambda_3$ and $M(\rho^a) = M(\rho^b)$ then $\vartheta_3(\rho^b) > \vartheta_3(\rho^a)$. \\

\noindent\textbf{CASE 2: If $\lambda_3>\lambda_2 :$} 
\begin{eqnarray}
M(\rho^a) &&= 1 - 4 (\lambda_1\lambda_4 + \lambda_2\lambda_3)^2 + 4\lambda_0^2 (\lambda_3^2 - \lambda_2^2) {}\nonumber\\&&= \cos^2\theta = M(\rho^b)
\end{eqnarray}
\begin{enumerate}
    \item \textbf{if $\lambda_3 > \lambda_2 \geq \lambda_4 \Rightarrow \vartheta_3(\rho^a) = 4\lambda_0\lambda_2+1$} \\
    Replace the value of $\cos\theta$ in $\vartheta_3(\rho^b)$ 
    \begin{align*}
    &\vartheta_3(\rho^b) = 2\cos\theta +1 \\
    &= 1+ 2\sqrt{1 - 4(\lambda_1\lambda_4 + \lambda_2\lambda_3)^2 + 4\lambda_0^2 (\lambda_3^2 - \lambda_2^2)} \\
    & > 1 + 2\sqrt{ 4\lambda_0^2\lambda_3^2 } > 1 + 4\lambda_0\lambda_2 = \vartheta_3(\rho^a)
    \end{align*}
    \item \textbf{if $\lambda_4 > \lambda_2 \Rightarrow \vartheta_3(\rho^a) = 4\lambda_0\lambda_4+1$} \\ 
    Similarly, simplify the last term to the form:
    \begin{align*}
    &\vartheta_3(\rho^b) = 2\cos\theta +1 \\
    &= 1+ 2\sqrt{1 - 4(\lambda_1\lambda_4 + \lambda_2\lambda_3)^2 + 4\lambda_0^2 (\lambda_3^2 - \lambda_2^2)} \\
    & > 1 + 2\sqrt{ 4\lambda_0^2\lambda_4^2} = 1 + 4\lambda_0\lambda_4 = \vartheta_3(\rho^a)
    \end{align*}
\end{enumerate}
\noindent Thus, if $\lambda_3>\lambda_2$ and $M(\rho^a) = M(\rho^b)$ then $\vartheta_3(\rho^b) > \vartheta_3(\rho^a)$. Thus, for all arbitrary states belonging to the GHZ class as defined above, the theorem is true, achieving equality only if $\lambda_2=\lambda_3=0$. $\blacksquare$
\end{proof}

\begin{figure}
    \centering
    \includegraphics[width=0.4\textwidth]{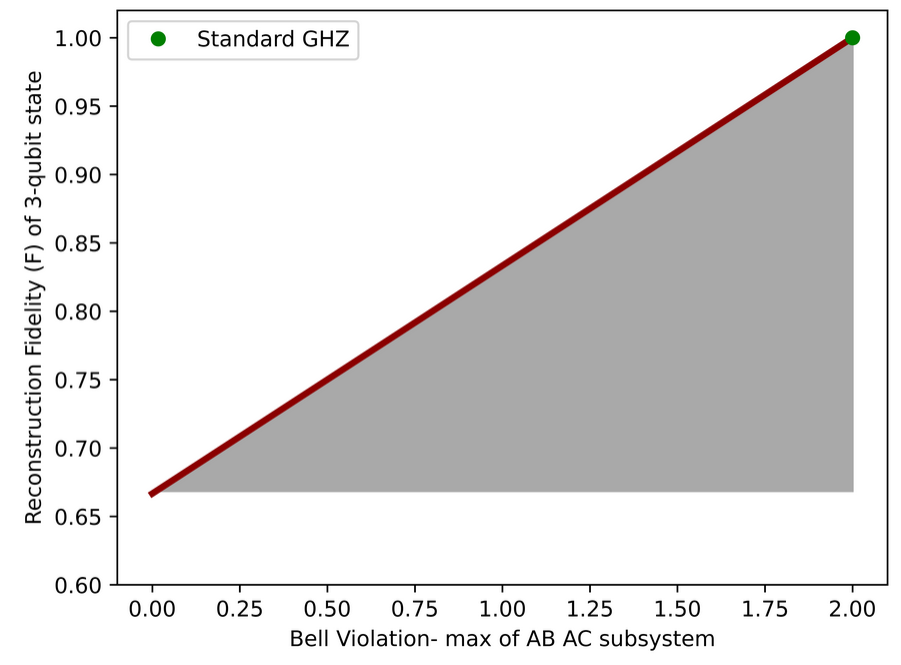}
    \caption{Reconstruction fidelity vs max of  Bell's CHSH values: The reconstruction fidelity of three-qubit pure states \cite{acin2001three,acin2001classification} is plotted against the maximum of Bell's CHSH values for both dealer--reconstructor and dealer--assistant channels. The red line represents the states taking the maximum value for the reconstruction fidelity. The green dot being the standard GHZ state is depicted as an example in the figure}
    \label{fig:rf-vs-bell}
\end{figure}

\noindent This is also an important result, proving that the maximum value of $M$ doesn't exceed $1$. In other words, the maximum of Bell's CHSH values of both the dealer--reconstructor and dealer--assistant channels is less than equal to $2$ (i.e $\mathcal{S}(\rho) \leq 2)$ in all cases of a secret shareable state. This means  that no relevant bipartitions of a state agreeing to the aforementioned conditions to qualify for faithful secret sharing violates Bell's inequality. In the Figure (\ref{fig:rf-vs-bell}) we numerically show that this is true for much larger class of states like Generalized GHZ class (Acin states \cite{acin2001classification,acin2001three}) with boundary being given by the red colour line. The standard GHZ state is one example in the boundary state, represented by the green point in the Figure (\ref{fig:rf-vs-bell}). \\ 

\noindent \textbf{Note:} Refer to the Appendix for complete calculations. \\ 

\section{Conclusion}

\noindent In conclusion, this work  identifies those secret sharing resources for which  maximum reconstruction of the secret is possible. Our result is based on the set of  Acin states representing the class of all tripartite pure states. We prove the analytical results for Acin states with vanishing complex coefficient. We call this class as Maximal Secret Reconstructed (MSR) class of states. This class of states are obtained as a result of the bound that we establish on the reconstruction fidelity for a given value of the maximum of bipartite teleportation fidelities of the dealer--assistant and dealer--reconstructor channels. 
\noindent Additionally, our study also evaluates the suitability of a state for reliable secret sharing in light of the Bell-CHSH operator values. Remarkably, we find that all secret shareable states does not violate Bell's inequality when observed across both the dealer--reconstructor and dealer--assistant partitions bringing out a new mutual exclusivity. In the same spirit, we also find the the maximum reconstruction fidelity that can be achieved if the Bell-CHSH values in the dealer--reconstructor and dealer--assistant partitions are known. Our study has been fundamental in the spirit of realizing the capacity of secret sharing states in terms of reconstructing the  secret bringing out the limits of probable resource theory based on quantum secret sharing. It also brings out the mutually exclusiveness of both bipartite correlations with tripartite capacity of a quantum states in a general sense conforming to such previous insights in different context \cite{pandya2016complementarity}.  


\bibliographystyle{unsrt}
\bibliography{ref.bib}

\appendix

\section{Teleportation Fidelity Detailed Analysis}
\noindent Given the correlation matrices $R$ and $Q$ for the subsystems AC and AB, we can calculate the value of $\vartheta_2(\rho) = \max\{||R||,||Q||\}$. where,
\begin{eqnarray}
    &&||R|| = 2 \lambda_0 \lambda_2 + 2 \sqrt{\lambda_0^2 \lambda_2^2 + (\lambda_1\lambda_2 + \lambda_3 \lambda_4)^2} {}\nonumber\\&&
    + \sqrt{4 \lambda_0^2 \lambda_1^2 + (\lambda_0^2 - \lambda_1^2 + \lambda_4^2 +\lambda_2^2 - \lambda_3^2)^2} {}\nonumber\\&&
    ||Q|| = 2 \lambda_0 \lambda_3 + 2 \sqrt{\lambda_0^2 \lambda_3^2 + (\lambda_1\lambda_3 + \lambda_2 \lambda_4)^2} {}\nonumber\\&&
    + \sqrt{4 \lambda_0^2 \lambda_1^2 + (\lambda_0^2 - \lambda_1^2 + \lambda_4^2 +\lambda_3^2 - \lambda_2^2)^2}
\end{eqnarray}
Analysing these values in the cases we need to compare with reconstruction fidelity, we can rewrite the values.\\ 

\noindent\textbf{CASE 1: if $\lambda_2>\lambda_3 :$} \\
\begin{eqnarray}
    &&\vartheta_2(\rho) = 2 \lambda_0 \lambda_2 + 2 \sqrt{\lambda_0^2 \lambda_2^2 + (\lambda_1\lambda_2 + \lambda_3 \lambda_4)^2} {}\nonumber\\&&
    + \sqrt{4 \lambda_0^2 \lambda_1^2 + (\lambda_0^2 - \lambda_1^2 + \lambda_4^2 +\lambda_2^2 - \lambda_3^2)^2} {}\nonumber\\&&
    = 2\lambda_0\lambda_2 + 2\sqrt{2\lambda_0^2\lambda_2^2 + (\lambda_1\lambda_2 + \lambda_3\lambda_4)} {}\nonumber\\&&
    + \sqrt{4\lambda_0^2\lambda_4^2 + 4\lambda_0^2(\lambda_2^2-\lambda_3^2) + (\lambda_2^2 + \lambda_4^2 - \lambda_0^2 - \lambda_1^2 - \lambda_3^2)^2}{}\nonumber\\&&
\end{eqnarray}
\noindent\textbf{CASE 2: if $\lambda_3>\lambda_2 :$} \\
\begin{eqnarray}
    &&\vartheta_2(\rho)= 2 \lambda_0 \lambda_3 + 2 \sqrt{\lambda_0^2 \lambda_3^2 + (\lambda_1\lambda_3 + \lambda_2 \lambda_4)^2} {}\nonumber\\&&
    + \sqrt{4 \lambda_0^2 \lambda_1^2 + (\lambda_0^2 - \lambda_1^2 + \lambda_4^2 +\lambda_3^2 - \lambda_2^2)^2} {}\nonumber\\&&
    = 2\lambda_0\lambda_2 + 2\sqrt{2\lambda_0^2\lambda_2^2 + (\lambda_1\lambda_2 + \lambda_3\lambda_4)} {}\nonumber\\&&
    + \sqrt{4\lambda_0^2\lambda_4^2 + 4\lambda_0^2(\lambda_3^2-\lambda_2^2) + (\lambda_0^2 + \lambda_1^2 + \lambda_2^2 - \lambda_3^2 - \lambda_4^2)^2}{}\nonumber\\&&
\end{eqnarray}
Using these simplifications, it is easy to see that 
\begin{equation}
    \vartheta_2(\rho) > 2\lambda_0\lambda_2 + 2\lambda_0\lambda_4
\end{equation}
\section{Bell Violation Detailed Analysis}
\noindent For subsystems AC and AB, we have the correlation matrices $R$ and $Q$. The eigenvalues of $R^TR$ and $Q^TQ$ can be calculated, respectively as:
\begin{align*}
[\hspace{0.2cm} & 4\lambda_0^2\lambda_2^2, \\
& 4 \lambda_0^2 \lambda_2^2 + 4(\lambda_1\lambda_2 + \lambda_3 \lambda_4)^2, \\
&4 \lambda_0^2 \lambda_2^2 + 4\lambda_0^2(\lambda_4^2 - \lambda_3^2) + (\lambda_0^2 + \lambda_1^2 +\lambda_3^2 - \lambda_2^2 - \lambda_4^2)^2 \\
]
\end{align*}
and 
\begin{align*}
[\hspace{0.2cm} & 4\lambda_0^2\lambda_3^2, \\
& 4 \lambda_0^2 \lambda_3^2 + 4(\lambda_1\lambda_3 + \lambda_2 \lambda_4)^2, \\
&4 \lambda_0^2 \lambda_3^2 + 4\lambda_0^2(\lambda_4^2 - \lambda_2^2) + (\lambda_0^2 + \lambda_1^2 +\lambda_2^2 - \lambda_3^2 - \lambda_4^2)^2 \\
]
\end{align*}
\noindent Consequently, it is obvious that the largest two values are the last two in each case.
\begin{equation*}
    M = 1 - 4 (\lambda_1\lambda_4 + \lambda_2\lambda_3)^2 + 4\lambda_0^2 |\lambda_2^2 - \lambda_3^2|
\end{equation*}
\noindent We can simplify $M$ in multiple ways, to compare with reconstruction fidelity in each case.\\

\noindent\textbf{CASE 1: if $\lambda_2>\lambda_3 :$} 
\begin{eqnarray}
    M &&=  1 - 4(\lambda_1\lambda_4 + \lambda_2\lambda_3)^2 + 4\lambda_0^2 (\lambda_2^2 - \lambda_3^2){}\nonumber\\&&
    =  4\lambda_0^2\lambda_2^2 +  (\lambda_0^2 + \lambda_1^2 - \lambda_2^2 + \lambda_3^2 - \lambda_4^2)^2 {}\nonumber\\&&
     + 4\lambda_0^2(\lambda_2^2 - \lambda_3^2) + 4(\lambda_1\lambda_2 + \lambda_3\lambda_4)^2 + 4\lambda_0^2\lambda_4^2 {}\nonumber\\&&
    =  4\lambda_0^2\lambda_4^2 +  (\lambda_0^2 + \lambda_1^2 - \lambda_2^2 + \lambda_3^2 - \lambda_4^2)^2 {}\nonumber\\&&
     + 4\lambda_0^2(2\lambda_2^2 - \lambda_3^2) + 4(\lambda_1\lambda_2 + \lambda_3\lambda_4)^2
\end{eqnarray}\\

\noindent\textbf{CASE 2: if $\lambda_3>\lambda_2 :$} 
\begin{eqnarray}
    M &&= 1 - 4(\lambda_1\lambda_4 + \lambda_2\lambda_3)^2 + 4\lambda_0^2 (\lambda_3^2 - \lambda_2^2) {}\nonumber\\&&
    =  4\lambda_0^2\lambda_3^2 +  (\lambda_0^2 + \lambda_1^2 + \lambda_2^2 - \lambda_3^2 - \lambda_4^2)^2 {}\nonumber\\&&
     + 4\lambda_0^2(\lambda_3^2 - \lambda_2^2) + 4(\lambda_1\lambda_3 + \lambda_2\lambda_4)^2 + 4\lambda_0^2\lambda_4^2 {}\nonumber\\&&
    =  4\lambda_0^2\lambda_4^2 +  (\lambda_0^2 + \lambda_1^2 + \lambda_2^2 - \lambda_3^2 - \lambda_4^2)^2 {}\nonumber\\&&
     + 4\lambda_0^2(2\lambda_3^2 - \lambda_2^2) + 4(\lambda_1\lambda_3 + \lambda_2\lambda_4)^2
\end{eqnarray}
It is now obvious that 
\begin{equation}
    M > 4\lambda_0^2\lambda_2^2 + 4\lambda_0^2\lambda_4^2
\end{equation}
\end{document}